\documentclass[journal,twoside,web]{ieeecolor}
\usepackage{tmi}
\usepackage{cite}
\usepackage{url}
\usepackage{lineno,hyperref}
\usepackage{amsmath,amssymb,amsfonts}
\usepackage[ruled]{algorithm2e} 
\usepackage{graphicx}
\usepackage{textcomp}
\usepackage{booktabs}
\usepackage{makecell}
\usepackage{multirow}
\usepackage[table,xcdraw]{xcolor}
\def\BibTeX{{\rm B\kern-.05em{\sc i\kern-.025em b}\kern-.08em
    T\kern-.1667em\lower.7ex\hbox{E}\kern-.125emX}}
\begin{document}
\title{Unveiling Incomplete Modality Brain Tumor Segmentation: Leveraging Masked Predicted Auto-Encoder and Divergence Learning}
\author{Zhongao Sun, Jiameng Li, Yuhan Wang, Jiarong Cheng, Qing Zhou, and Chun Li, \IEEEmembership{Member, IEEE}
\thanks{This research was supported by the University-Level Research and Innovation Project of Shenzhen MSU-BIT University.} 
\thanks{Z. Sun, J. Li, Y. Wang, and C. Li are with Shenzhen MSU-BIT University, Shenzhen, 518172, China. E-mail:(sunzhongao0224@gmail.com; leejiameng2019@gmail.com; smbu\_wyh@outlook.com; LF\_cyan27@outlook.com).}
\thanks{*Corresponding author: Chun Li (E-mail: lichun2020@smbu.edu.cn).}
}

\maketitle

\begin{abstract}
Brain tumor segmentation remains a significant challenge, particularly in the context of multi-modal magnetic resonance imaging (MRI) where missing modality images are common in clinical settings, leading to reduced segmentation accuracy. To address this issue, we propose a novel strategy, which is called masked predicted pre-training, enabling robust feature learning from incomplete modality data. Additionally, in the fine-tuning phase, we utilize a knowledge distillation technique to align features between complete and missing modality data, simultaneously enhancing model robustness. Notably, we leverage the Hölder pseudo-divergence instead of the Kullback–Leibler divergence (KLD) for distillation loss, offering improve mathematical interpretability and properties. Extensive experiments on the BRATS2018 and BRATS2020 datasets demonstrate significant performance enhancements compared to existing state-of-the-art methods.
\end{abstract}


\section{Introduction}
\label{sec:introduction}
\begin{figure}[t!]
    \centering
    \includegraphics[width=0.5\textwidth]{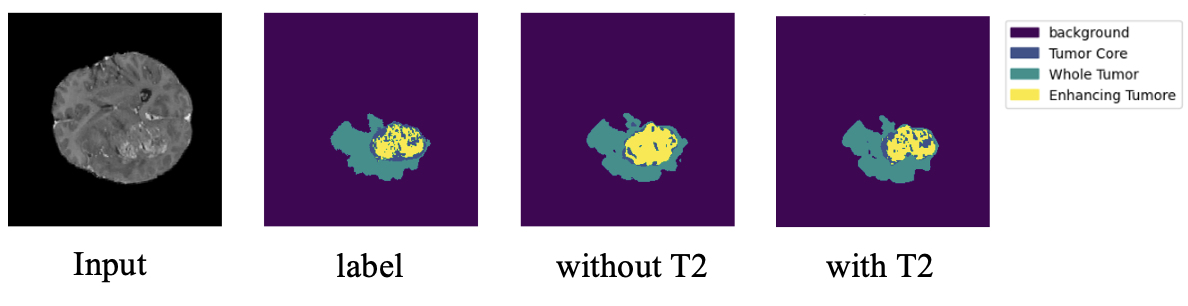}
    \caption{Influence on enhancing tumor when missing T2 modality}
    \label{fig:T2}
\end{figure}

\IEEEPARstart{T}he segmentation of brain tumors using magnetic resonance imaging (MRI) is indispensable for clinical evaluations and diagnoses, offering detailed insights into brain anatomy and pathology to aid in precise treatment planning and disease progression monitoring \cite{brain_survial}. Typically, four MRI modalities - T1-weighted (T1), contrast-enhanced T1-weighted (T1c), T2-weighted (T2), and T2 fluid attenuation inversion recovery (FLAIR) - are employed to distinguish various brain tissues. The impact of missing T2 modality on tumor enhancement is depicted in Fig. \ref{fig:T2}, highlighting its significance within the context of image processing for tumor segmentation. Integrating multimodal imaging techniques for brain tumor segmentation significantly enhances segmentation precision. While existing methods achieve high accuracy by utilizing all four modalities as input \cite{nnunet,Contribution2017, SwinUNETR,multimoseg,sslswin}, they are designed to accommodate all imaging modalities. In real-world scenarios, the absence of one or more imaging modalities is common due to data corruption, diverse scanning protocols, or patient conditions \cite{inresp,10.1007/978-3-030-59719-1_52}. Consequently, there is a pressing need for a resilient multimodal approach that offers flexibility and practicality for clinical applications, specifically addressing challenges associated with missing modalities.

Several methods have been proposed to address this issue, with modal generation-based approaches utilizing generative adversarial networks (GAN) \cite{gan} or Gaussian process prior variational autoencoder (MGP-VAE) \cite{MGP-VAE} to generate missing modalities from available ones. However, these methods may underperform when only a single modality is available. Alternatively, modality-specific methods map each modality into a latent space and employ various fusion strategies to combine features. For instance, RFNet \cite{rfnet} aggregates multi-modal features from different regions adaptively to model modality and tumor region relations, yielding superior results. Ting and Liu \cite{Multimodal} utilize modality-specific encoders and a multimodal shared-weight decoder to learn modality-specific features. Despite the variety of fusion methods, inter-modal interaction remains lacking \cite{Qiu_2023_ICCV}. Additionally, knowledge distillation proves effective in enhancing the performance of missing modality networks by transferring knowledge from full modality networks. D$^2$-Net \cite{D2-Net} incorporates additional stages for modality and tumor-region disentanglement, along with a knowledge module, enabling feature extraction robustness even with missing modalities. Similarly, SMU-Net \cite{Azad2022SMUNetSM} employs distillation not only in prediction masks but also in latent feature space.

In recent advancements in computer vision, mask imaging modeling has emerged as a significant area of progress \cite{MAE,swinv2,SimMIM,sslswin,Swin_MAE}. Furthermore, the combination of mask imaging modeling and knowledge distillation has shown promising results \cite{dbot,bai2022masked}. M$^3$AE \cite{M3AE} pioneers the fusion of masked auto-encoder and knowledge distillation in multimodal representation learning with missing modalities, offering an efficient model applicable to all possible modal subsets. However, pre-training by sampling random subsets may lead to sub-optimal performance due to potential information loss between modalities. Ding et al. \cite{rfnet} observed distinct appearances in different modalities, such as Flair's sensitivity to background and T2's sensitivity to peritumoral edema (ED). Consequently, in the absence of specific modalities, models may need to focus on different regions. Drawing inspiration from predicting the next word \cite{beit}, we propose a novel pre-training strategy that reconstructs both missing modalities and masked patches to extract both intra and inter-modal information. Additionally, we replace traditional transformer blocks with 3D swin transformer blocks \cite{sslswin} and explore the potential of Hölder divergence in the fine-tuning stage to transfer knowledge from full-modal networks to missing-modal networks.

Overall, our contributions are threefold:
\begin{enumerate}
    \item We introduce a novel pre-training framework inspired by predicting the next word and mask imaging modeling, which predicts both masked patches and missing modalities. Our experiments demonstrate that this approach effectively extracts both intra-modal and inter-modal information, surpassing methods that solely reconstruct masked patches or missing modalities.

    \item  Additionally, we explore divergence learning in the knowledge distillation (KD) stage, comparing Hölder divergence with KL divergence in distilling segmentation logits.

     \item  By leveraging the proposed masked predicted auto-encoder and Hölder divergence, our model achieves state-of-the-art performance in handling missing modalities across most cases in the BraTS2018 benchmarks.
\end{enumerate}
The main notations used in this work is shown in Table \ref{tab-0}. 

\section{Related Works}
\subsection{Incomplete Multi-Modal Brain Tumor Segmentation}
Compared to standard brain tumor segmentation methods \cite{nnunet,Contribution2017, SwinUNETR,multimoseg, sslswin, p9, p10}, incomplete multimodal segmentation poses greater challenges but is more practical. In brain tumor segmentation evaluation, three crucial indicators - Whole Tumor (WT), Tumor Core (TC), and Enhancing Tumor (ET) - must be considered. The absence of one or more modalities can significantly reduce Dice results. For instance, missing modality T1c may result in over a 40\% decrease in the Enhancing Tumor index.

Several methods have been proposed to address this issue. Some utilize separate encoders for each modality to extract modality-specific features. For example, MMFormer \cite{mmFormer} employs separate convolutional encoders and additional transformer blocks to capture long-range dependencies across modalities. RFNet \cite{rfnet} introduces a region-aware fusion module to aggregate multimodal features. Ting and Liu \cite{Multimodal} utilize modality-specific encoders, a multimodal shared-weight decoder, and a missing-full complementary strategy to learn correlations between missing and full modalities. However, these methods often employ complex feature fusion techniques, which are challenging to understand and transfer to other domains. Furthermore, although fused after feature extraction, the use of modality-specific extractors inevitably lacks interaction information between modalities \cite{Qiu_2023_ICCV}.

Alternatively, some approaches aim to address all problems by employing dedicated models for each target missing situation. For example, Avrim and Mitchell \cite{Co-Training} attempt to distill knowledge from full-modal to missing-modal networks. And SMU-Net \cite{Azad2022SMUNetSM} proposes distilling knowledge from a multimodal teacher network to missing modal students at the latent space and network output levels. ProtoKD \cite{Wang2023} utilizes prototype learning combined with distillation to model various missing modalities, achieving promising results. These methods effectively mitigate the lack of several modalities, which is the most challenging aspect in practice.

Therefore, we propose utilizing separate models for each situation to address the absence of multiple modalities. This strategy avoids inference overhead and can lead to performance improvements.

\subsection{Masked Imaging Modeling}
Since the introduction of Devlin et al. \cite{bert}, a masked-based model, masking part of tokens and predicting the rest has emerged as a new pre-training paradigm in natural language processing. Inspired by this idea, Devlin et al. \cite{beit} pioneered the use of masked modeling in the computer vision field, achieving significant success. Subsequent studies such as SimMIM \cite{SimMIM} and MAE \cite{MAE} discovered that reconstructing RGB pixels from masked tokens during the pre-training stage can lead to state-of-the-art classification accuracy, even with pre-training solely on ImageNet-1k \cite{imagenet}. MAE adopts an encoder-decoder architecture with a larger encoder and a lightweight decoder using traditional transformer blocks. In contrast, SimMIM achieves similar accuracy using a one-layer prediction head as the decoder.

To address dense prediction tasks like semantic image segmentation, swin transformer \cite{swin} and masked-based pretrained swin transformer v2 \cite{swinv2} are developed. By incorporating a shifted-window mechanism and hierarchical structure, swin transformer can better capture local information, leading to state-of-the-art results in semantic segmentation and object detection.

Despite the smaller scale of datasets in medical image analysis compared to general image datasets, the success of Swin MAE \cite{Swin_MAE} and self-supervised Swin Transformer for 3D medical images \cite{sslswin} demonstrates that the masked image modeling pre-training paradigm remains effective even with small datasets.

In this work, inspired by the success of masked modeling in predicting the next word, we apply this idea to missing modality pre-training, aiming to compel the model to learn intra-modality information. Additionally, we employ masked modeling strategies to better capture inter-modality features.

\subsection{Knowledge Distillation}
KD, a widely employ technique introduced by Hinton et al. \cite{Hinton2015DistillingTK}, was initially devised to transfer the knowledge of a teacher model to a smaller student model. Since its inception, numerous variations have been explored to enhance its efficacy. These include adjusting distillation targets to include intermediate features \cite{featuredis}, employing multi-stage distillation during pre-training \cite{dbot}, and enhancing distillation through regularization \cite{wang2023improving}, among others.

In the context of missing modality brain tumor segmentation, several studies have leveraged knowledge distillation to transfer knowledge from full-modal networks to missing modality networks. For instance, SMU-Net \cite{Azad2022SMUNetSM} incorporates a content and style-matching mechanism to distill informative features from both full and missing modalities in the latent space. MMCFormer \cite{karimijafarbigloo2023mmcformer} transfers modality-specific representations with the aid of auxiliary tokens. Moreover, M$^3$AE \cite{M3AE} distills shared semantics between heterogeneous missing-modal situations within a single network.

In our work, rather than solely focusing on distillation strategies or employing additional methods to improve performance, we investigate divergence learning with the aim of enhancing distillation using Hölder divergence. Specifically, we explore Hölder divergence in distilling the class probability distribution of each pixel and compare it with KLD.
\newtheorem{definition}{\bf{Definition}}
 KLD \cite{p1}, a fundamental metric used to assess the similarity between probability distributions. Which exists in two distinct forms: discrete and continuous scenarios, which can be defined as Definition \ref{def_1}:
\begin{definition} 
	\label{def_1}
	(\textbf{Kullback-Leibler Divergence \cite{p1}}) Consider a random variable $X$ with potential outcomes in the set $\Omega$, and let $P$ and $Q$ denote two probability distributions defined on $X$. 
	\begin{enumerate}
		\item The KLD of $P$ from $Q$ for a discrete random variable $X$ is formally defined as:
		\begin{equation}
			\label{equ_1}
			\operatorname{KL}[P\|Q]=\sum_{x\in\Omega}p(x)\cdot\log\frac{p(x)}{q(x)},
		\end{equation}
		where $p(x)$ and $q(x)$ represent the probability mass functions of $P$ and $Q$, respectively.
		
		\item The KLD of $P$ from $Q$ for a continuous random variable $X$ is formally defined as:
		\begin{equation}
			\label{equ_2}
			\operatorname{KL}[P||Q]=\int_{\Omega}p(x)\cdot\log\frac{p(x)}{q(x)}\operatorname{d}x,
		\end{equation}
		where $p(x)$ and $q(x)$ represent the probability density functions of $P$ and $Q$, respectively.
	\end{enumerate}
\end{definition}

\section{Methodology}
\begin{table}[t]
	\centering
	\caption{Main Notations Used in This Work.}
	\setlength{\tabcolsep}{3pt}
	\begin{tabular}{p{3cm}<{\raggedright}p{5cm}<{\raggedright}} 
		\toprule[1.5pt]
		Notation&Definition\\
		\midrule[0.5pt]
		${D_{KL}}(.||.)$& KL divergence \\
		$\alpha, \beta$ &  The conjugate exponents of Hölder \\
		$D_\alpha^H(p(x):q(x))$ & The Hölder pseudo-divergence of $p(x)$ and $q(x)$\\

		$\left\{ {\left\{ {{\rm X}_n^m} \right\}_{m = 1}^M,{Y_n}} \right\}_{n = 1}^N$& The $n$ samples with $M$ modalities each, and the labels corresponding to the $n$ samples, respectively\\
		\bottomrule[1.5pt]
	\end{tabular}
	\label{tab-0}
\end{table}	
\subsection{Perliminaries}
\subsubsection{Knowledge Distillation for Segmentation}:
In contrast to traditional image classification, segmentation involves assigning an individual category label to each pixel from $N$ category species. Let the input to the network be denoted as $F \in \mathbb{R}^{C\times D\times H \times W}$, where $C$, $D$, $H$, and $W$ represent the number of channels, depth, height, and width, respectively. The segmentation network transforms $F$ into a categorical logit map $\mathbf{S} \in \mathbb{R}^{C\times D\times H \times W}$. The loss function for the segmentation task aims to train each pixel with its ground-truth label using the soft Dice loss \cite{diceloss}:
\begin{equation}
\label{equ_3}
\mathcal{L}(G, Y)=1-\frac{2}{J} \sum_{j=1}^J \frac{\sum_{i=1}^I G_{i, j} Y_{i, j}}{\sum_{i=1}^I G_{i, j}^2+\sum_{i=1}^I Y_{i, j}^2},
\end{equation}
where $I$ denotes the number of voxels, while $J$ represents the number of classes. Here, $Y_{i, j}$ and $G_{i, j}$ denote the probability of class $j$ at voxel $i$ in the output and one-hot encoded ground truth, respectively.

Inspired by Hinton's KD approach \cite{Hinton2015DistillingTK}, a straightforward method is to align the class probability distribution of each pixel from the student to that of the teacher. The formulation of $L_{k d}$ is as follows:
\begin{equation}
\label{equ_4}
\small
\frac{1}{D \times H \times W} \sum_{d=1}^D \sum_{h=1}^H \sum_{w=1}^W K L\left(\sigma\left(\frac{\mathbf{S}_{d, h, w}^s}{\tau}\right) \| \sigma\left(\frac{\mathbf{S}_{d, h, w}^t}{\tau}\right)\right),
\end{equation}
where $\sigma(\frac{\mathbf{S}_{d, h, w}^t}{\tau})$ and $\sigma(\frac{\mathbf{S}_{d, h, w}^s}{\tau})$ represent the soft class probabilities of the $(d,h,w)$-th pixel produced from the student and teacher, respectively. KL denotes the Kullback-Leibler divergence, and $\tau$ is a temperature. 

In this work, we explore the relationship between KL divergence and Hölder divergence, aiming to enhance the performance of knowledge distillation.

\subsubsection{3D Swin Transformer Block}

We employ the Swin-UNet \cite{SwinUNETR} as our backbone model, featuring a U-Net \cite{unet}-shaped architecture equipped with swin transformer encoder blocks and a CNN decoder. Given an input to the network denoted as $F \in \mathbb{R}^{C\times D\times H \times W } $, we initially utilize a 3D convolutional block with a consistent kernel size and stride size $P$, generating an output channel size of $S$. This convolutional block aims to patchify the 3D brain tumor image into non-overlapping tokens. Subsequently, the dimensions become $S\times [\frac{D}{P}] \times [\frac{G}{P}] \times [\frac{W}{P}]$, and the input is forwarded to the swin transformer block, where the output is computed as follows:
\begin{equation}
\label{equ_5}
    \begin{cases}\hat{z}^{\prime}=\text{W}-\text{MSA}\left(\text{LN}\left(z^{l-1}\right)\right)+z^{l-1},\\z^{\prime}=\text{MLP}\Big(\text{LN}\Big(\hat{z}^{l}\Big)\Big)+\hat{z}^{l},\\\hat{z}^{l+1}=\text{SW- MSA}\Big(\text{LN}\Big(z^{l}\Big)\Big)+z^{l}\text{,}\\z^{l+1}=\text{MLP}\Big(\text{LN}\left(\hat{z}^{l+1}\right)\Big)+\hat{z}^{l+1},\end{cases}
\end{equation}
where MLP and LN denote layer normalization and multi-layer perceptron (MLP), respectively.

To mitigate the quadratic complexity problem associated with self-attention in the original vision transformer \cite{vit}, we adopt a window-based multi-head self-attention (W-MSA) approach to concentrate attention within localized windows. Furthermore, shifted window multi-head self-attention (SW-MSA) is employed to improve connectivity between these windows. For 3D imaging, we modify the 3D cyclic-shifting technique \cite{swin} to meet our specific needs.

Informed by the Swin UNet \cite{SwinUNETR}, we set the patch size $P$ and output channel size $S$ to 2 and 48, respectively. Additionally, the encoder consists of 4 stages, with each stage comprising 2 transformer blocks. To reduce the resolution of feature representations while maintaining the model's hierarchical structure, a patch merging layer is integrated. Below is a concise summary of our model configuration present in Table \ref{tab:my_label}.

\renewcommand\arraystretch{1.0}
\begin{table}
     \setlength{\belowdisplayskip}{0pt}
     \setlength{\abovedisplayskip}{0pt}
     \setlength{\abovecaptionskip}{0pt}
     \centering
     \scriptsize
     \caption{A succinct configuration overview of our model.}
     \setlength{\tabcolsep}{14pt}
    \begin{tabular}{lll}
        \toprule[2pt] 
        Embed Dimension & Feature Size & \makecell{Number of  Blocks} \\
        \midrule[1.0pt]
        768& 48 & $[2,2,2,2]$ \\
        \midrule[1.0pt]
        \makecell{Window Size} & \makecell{Number of Heads} & Parameters \\
        \midrule[1.0pt] 
        $[7,7,7]$ & $[3,6,12,24]$ & $61.98 \mathrm{M}$ \\
        \bottomrule[2pt]
        \end{tabular}
    \label{tab:my_label}
\end{table}

\subsection{Self-Supervised Learning via Predicting Missing Modality}

Our pre-training strategy is specifically modified for the case of missing modalities, that the prediction targets not only the masked patches in visible modalities but also the  missing modalities. Our pretraining framework consists of 3 major components:

  \begin{enumerate}
      \item \textit{Encoder and Decoder Architecture}: The encoder is employed to acquire latent representations from the input 3D images, and subsequently, the pre-trained Encoder is loaded for downstream tasks, such as segmentation. In comparison to the conventional vision transformer block, the swin transformer block \cite{swin,SwinUNETR} proves more suitable for dense prediction tasks. On the other hand, the decoder is tasked with reconstructing the masked tokens and should be considerably lighter than the encoder \cite{SimMIM}. Following prior research \cite{sslswin,wang2023SwinMM}, the extracted representations are then fed into a straightforward CNN-based decoder.

    \item \textit{Masking Strategy}: When present with a 3D brain tumor image that lacks one or several modalities, the masking strategy determines the extent of information provided to the Encoder. This strategy is closely tied to the complexity of the pre-training task. Naturally, an increase in the number of missing modalities corresponds to greater difficulty in the prediction task. Therefore, the mask ratio should be adjusted downward to accommodate this heightened challenge. We delve into this scenario further in the subsequent section.
    
    \item \textit{Prediction Sarget}: The cornerstone of our strategy lies in an innovative approach to handling missing modalities. While prior studies predominantly focused on reconstructing only the masked areas \cite{sslswin, wang2023SwinMM, MAE}, typically aiming to restore the raw pixels or specific image features, we have taken a novel leap forward. Drawing inspiration from predictive techniques utilized in natural language processing \cite{bert}, such as predicting the next word in a sequence, we have extended this concept to address our unique challenge of missing modalities. We propose a task where the prediction of the absent modalities becomes an additional objective. This novel task is subsequently benchmarked against traditional reconstruction methods in the ensuing sections, where we conduct a comprehensive analysis and comparison of the objectives and their efficacy.
  \end{enumerate}

In the forthcoming subsections, we elucidate the \textit{Masking Strategy} and \textit{Prediction Target} of our model. These configurations will undergo a systematic examination to assess their efficacy. By strategically combining fundamental designs from each constituent part, we successfully cultivate robust representation learning capabilities for addressing missing modality situations.

\subsubsection{Masking Strategy}:
As mentioned earlier, when more modalities are missing, the mask ratio $p$ should be reduced accordingly. Therefore, we hypothesize a linear mapping relationship between them, formulated as follows:
\begin{equation}
\label{equ_6}
    p = k \times m + b,
\end{equation}
where $p$ denotes the mask ratio, $m$ represents the number of missing modalities, and $k$ and $b$ denote the parameters. Our experiments encompassed scenarios with no missing modalities as well as instances where T1, T1c, and T2 modalities are missing. Notably, we observe that in the absence of any missing modalities, the optimal mask ratio is $0.75$, yielding the best results. Conversely, with all three modalities missing, the best performance was achieved when the mask ratio is set to $0.5$. These findings are presented in the accompanying table. Additionally, inspired by SimMIM \cite{SimMIM}, we adopt a learnable mask token vector to replace each mask patch, which is then jointly input into the Encoder. It's important to note that the mask is applied across the channel dimension, ensuring consistency across all modalities.

\subsubsection{Prediction Target}:
We consider using the original pixel values as the target of reconstruction. The reconstruction loss can be written in the following form:
\begin{equation}
\label{equ_7}
    \mathcal{L}_{rec} = ||(x - x_{rec}) ||_{l1} \times mask,
\end{equation}
where $x$ is the input of the model, $x_{rec}$ is the reconstruction of the model, multiplying the $mask$ aims to focus only on the obscured area. When considering the case of simultaneously predicting missing modalities, the loss function can be modified as follows:
\begin{equation}
\label{equ_8}
    \mathcal{L}_{rec} = ||(concat(x_{visual},x_{missing}) - x_{rec}||_{l1} \times mask,
\end{equation}
where $x_{visual}$ represents the visual modalities of the input, while $x_{missing}$ denotes the missing modalities, respectively. And $x_{rec}$ stands for the reconstructed full modality image. This approach enables the model to simultaneously acquire the capability to extract both intra-modal and inter-modal information during the pre-training phase.

Furthermore, we compare different prediction targets, including scenarios with no mask, exclusively predicting the missing modalities, and solely predicting the missing areas. Our experimental results, present in the accompanying table, demonstrate that our proposed method of simultaneously predicting both the missing modality and region enhances the model's efficacy in downstream tasks.

\begin{figure*}[ht!]
    \centering
    \includegraphics[width=\textwidth]{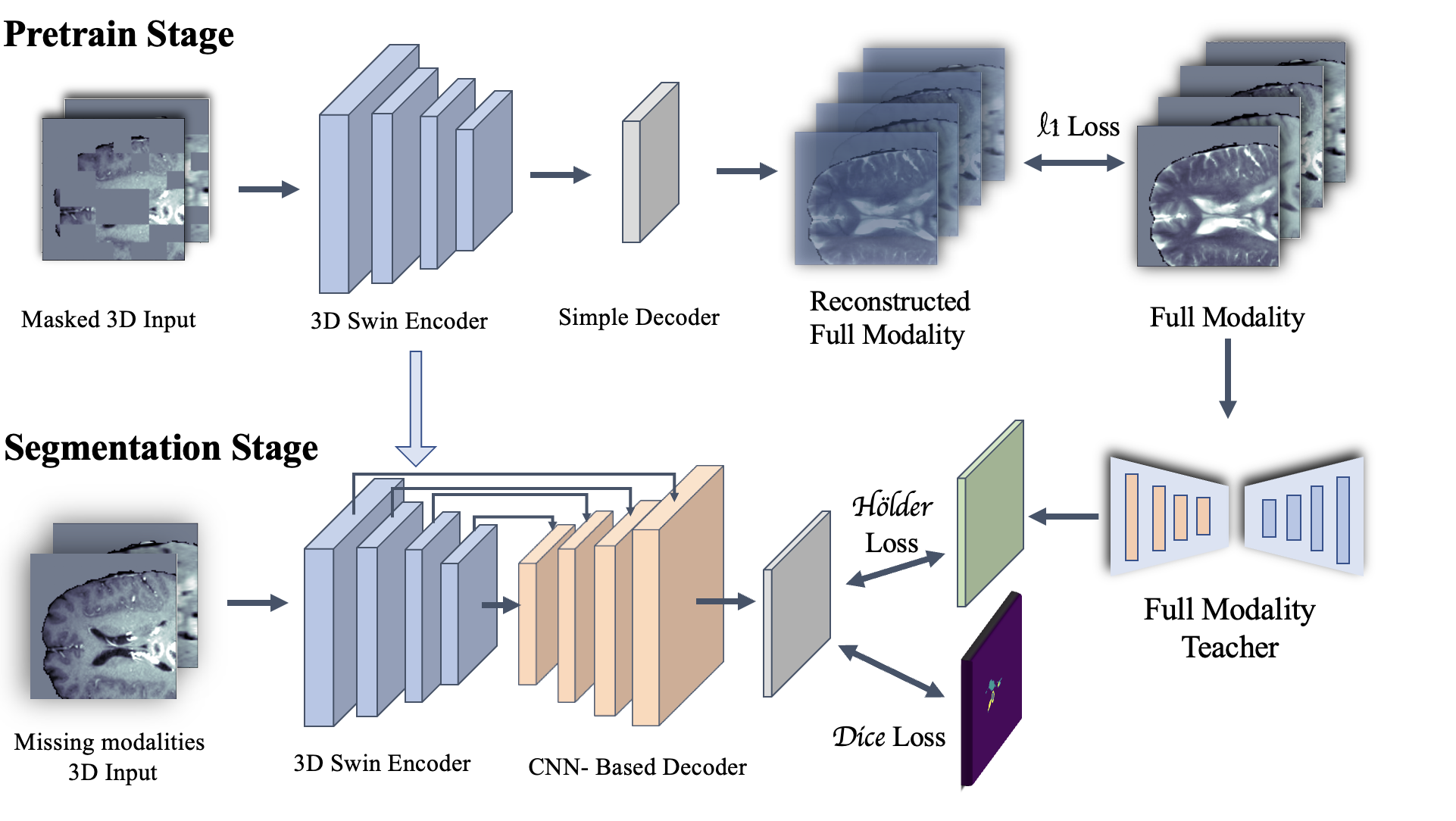}
    \caption{The framework of our unveiling incomplete modality brain tumor segmentation: leveraging masked predicted auto-encoder and divergence Learning.}
    \label{fig:WSI}
\end{figure*}

\subsection{Exploring Incomplete Modality Brain Tumor Segmentation via Hölder Divergence Analysis}

\subsubsection{Related Essential Definitions of Hölder Divergence}
The Hölder Divergence, initially introduced in 2014 \cite{Holder47,Holder48,Holder49}, shares similarities with the concept of KLD \cite{p1}. Hölder Divergence leverages the notion of strictness in inequalities to establish a framework for measuring dissimilarity. When present with an inequality in the form of $lhs \leq rhs$, where $lhs$ and $rhs$ denote the left and right sides of the inequality respectively, we evaluate the strictness, denote as $\Delta = rhs - lhs$. In situations where $lhs>0$, we can define the following disparity:
\begin{equation}
\label{equ_9}
    D=\log \frac{r h s}{l h s}=-\log \frac{l h s}{r h s} \geq 0,
\end{equation}

In the realm of two-parameter inequalities denote as $\operatorname{lhs}(p, q) \leq \operatorname{rhs}(p, q)$ where $p \neq q$, consistency in the condition $\operatorname{lhs}(p, q) < \operatorname{rhs}(p, q)$ characterizes the inequality as proper. Conversely, when $p=q$ and $\operatorname{lhs}(p, q) = \operatorname{rhs}(p, q)$ holds true, the inequality is termed tight. Moreover, a proper inequality may signify a proper variation. Specifically, in scenarios where $q=p$ and $D(p, q)=0$, a correct adjustment is indicated. Inequalities deviating from the category of proper inequalities are deemed ill-posed inequalities. Subsequently, pseudo-divergence is introduced based on ill-posed inequalities. Hölder divergence is defined by leveraging the Hölder inequality, which states that for two positive real functions $p(x)$ and $q(x)$ within the same probability space $\Omega$, the following relationship holds valid:
\begin{equation}
\label{equ_10}
    \int_\Omega  p (x)q(x)dx \le {\left( {\int_\Omega  p {{(x)}^\alpha }dx} \right)^{\frac{1}{\alpha }}}{\left( {\int_\Omega  q {{(x)}^\beta }dx} \right)^{\frac{1}{\beta }}},
\end{equation}
where $\alpha$ and $\beta$ satisfy $\alpha \beta>0$ and $\frac{1}{\alpha}+\frac{1}{\beta}=1$. They are referred to as a pair of Hölder conjugate exponents. For the Hölder conjugate exponents $\alpha$ and $\beta$, if the probability density functions $p(x) \in L^\alpha(\Omega, \mu)$ and $q(x) \in L^\beta(\Omega, \mu)$. 

According to \cite{Holder49}, the formulation of Hölder Divergence not only encompasses the notion of inequality tightness but also encapsulates two distinct cases: Hölder Statistical Pseudo-Divergence (HPD) and Proper Hölder Divergence (PHD). Below, based on Definition \ref{def_1}, we present their respective definitions:

\begin{definition} 
	\label{def_3}
	(\textbf{Hölder Statistical Pseudo-Divergence, HPD \cite{Holder49}}) HPD pertains to the conjugate exponents $\alpha$ and $\beta$, where $\alpha \beta>0$. In the context of two densities, $p(x) \in {L^\alpha }\left( {\Omega,\nu } \right)$ and $q(x) \in {L^\beta }\left( {\Omega ,\nu } \right)$, both of which belong to positive measures absolutely continuous with respect to $\nu$, HPD is defined as the logarithmic ratio gap, as follows:
	\begin{equation}
		\label{equ_8}
		D_{\alpha}^{H}(p(x):q(x))=-\log\left(\frac{\int_{\Omega}p(x)q(x)\mathrm{d}x}{\left(\int_{\Omega}p(x)^{\alpha}\mathrm{d}x\right)^{\frac1\alpha}\left(\int_{\Omega}q(x)^{\beta}\mathrm{d}x\right)^{\frac1\beta}}\right).
	\end{equation}
	When $0<\alpha<1$ and $\beta  = \bar \alpha  = \frac{\alpha }{{\alpha  - 1}} < 0$ or $\alpha<0$ and $0<\beta<1$, the reverse HPD is defined by:
	\begin{equation}
		\label{equ_9}
		\begin{aligned}D_\alpha^\mathrm{H}(p(x):q(x))&=\log\left(\frac{\int_{\Omega}p(x)q(x)\mathrm{d}x}{\left(\int_{{\Omega}}p(x)^\alpha\mathrm{d}x\right)^{\frac1\alpha}\left(\int_{{\Omega}}q(x)^\beta\mathrm{d}x\right)^{\frac1\beta}}\right).\end{aligned}
	\end{equation}
\end{definition}

\begin{definition} 
	\label{def_4}
	(\textbf{Proper Hölder Divergence, PHD \cite{Holder49}}) The proper Hölder divergence between two densities $p(x)$ and $q(x)$ is defined for conjugate exponents $\alpha, \beta>0$, and $\gamma>0$, where $D_{\alpha ,\gamma }^{\rm{H}}(p(x):q(x)) = D_\alpha ^{\rm{H}}(p{(x)^{\gamma /\alpha }}:q{(x)^{\gamma /\beta }})$ is:
	\begin{equation}
		\label{equ_10}
          D_{\alpha ,\gamma }^{\rm{H}} =  - \log \left( {\frac{{\int_\Omega  p {{(x)}^{\gamma /\alpha }}q{{(x)}^{\gamma /\beta }}{\rm{d}}x}}{{{{(\int_\Omega  p {{(x)}^\gamma }{\rm{d}}x)}^{1/\alpha }}{{(\int_\Omega  q {{(x)}^\gamma }{\rm{d}}x)}^{1/\beta }}}}} \right),
	\end{equation}
	where by definition, $D_{\alpha, \gamma }^{\rm{H}}(p(x):q(x))$ constitutes a two-parameter family of statistical dissimilarity measures.
\end{definition}

\subsubsection{Utilize Hölder divergence in Knowledge Distillation of Segmentation}
The segmentation problem can be conceptualized as a pixel-level classification task. In our scenario, there are a total of four classes: 1) background, 2) whole tumor, 3) tumor core, and 4) enhancing tumor. As mentioned earlier, in cases of missing modalities, the deletion of different modalities results in a significant decrease in accuracy across various classes. KLD typically accentuates certain categories with prominent features while disregarding others, primarily due to information loss from the missing modalities \cite{holder_zhangyan}. Conversely, Hölder divergence exhibits distinct characteristics, featuring a consistently increasing curve and the ability to distinguish all categories, even those with notable features \cite{holder_zhangyan}.

Moreover, Hölder divergences encompass both the Cauchy-Schwarz divergence and the one-parameter family of skew Bhattacharyya divergences, making them particularly noteworthy \cite{Holder48,Holder49}. In situations where the natural parameter space forms a cone or an affine, Hölder divergences allow for closed-form expressions between distributions belonging to the same exponential families. This proves advantageous when examining exponential families with conic or affine natural parameter spaces, such as multinomials or multivariate normals \cite{Holder48,Holder49}.

In accordance with \cite{Holder49}, the mathematical properties of Hölder divergence have been analyzed, showcasing its exceptional mathematical characteristics. Consequently, Hölder divergence has found diverse applications \cite{p2,p3,p4,p5,p6}.

In this work, Hölder divergence is employed as a substitute for KLD. We opt not to employ any specialized distillation strategy but rather used simple pixel-wise knowledge distillation. This approach aligns the missing modality student model with the full modality teacher model by minimizing Hölder divergence. The loss function is defined as follows:
\begin{equation}
\label{equ_11}
\small
\frac{1}{D \times H \times W} \sum_{d=1}^D \sum_{h=1}^H \sum_{w=1}^W D_\alpha^H  
\left(\sigma\left(\frac{\mathbf{S}_{d, h, w}^s}{\tau}\right) \| \sigma\left(\frac{\mathbf{S}_{d, h, w}^t}{\tau}\right)\right),
\end{equation}
where $\sigma(\frac{\mathbf{S}_{d, h, w}^t}{\tau})$ and $\sigma(\frac{\mathbf{S}_{d, h, w}^s}{\tau})$ represent the soft class probabilities of the $(d,h,w)$-th pixel produced from the student and teacher respectively. $D_\alpha^H$ denotes the Hölder divergence, and $\tau$ and $\alpha$ is temperature coefficient and Hölder conjugate exponents, respectively. The algorithm proposed in this study is presented in Algorithm \ref{alg:spl}. And the framework of our proposed model is illustrated in Fig. \ref{fig:WSI}.
\begin{algorithm}[t]
	\caption{\small Unveiling Incomplete Modality Brain Tumor Segmentation: Leveraging Masked Predicted Auto-Encoder and Divergence Learning.}
	\label{alg:spl}
	\DontPrintSemicolon
	\small
	\tcp*[f]{\textbf{*Pre-training*}}\\
	\textbf{Input:} Multi-Modality Dataset: $D = \left\{ {\left\{ {{\rm X}_n^m} \right\}_{m = 1}^M,{Y_n}} \right\}_{n = 1}^N$;\\
	\textbf{initialization:} Initialize the parameters of the model.\\
	\While{not converged}  
	{
        $x_{full}$ $\leftarrow$ full modality batch data
        \\
        $x_{visible}$ $\leftarrow$ visible modality batch data 
        \\
        $x_{masked}$, $mask$ $\leftarrow$ mask\_rand\_patch $(x_{visible}, mask_{ratio})$
        \\
        $x_{rec}$ $\leftarrow$ model $(x_{masked})$
        \\
        loss $\leftarrow$ MSE $(x_{rec}$, $x_{full}$, $mask)$
		
	}
	{\bfseries Output:} model parameters.\\
	\tcp*[f]{\textbf{*Fine-tuning*}}\\
	\textbf{initialization:} Transfer the parameters of the pre-trained Encoder.\\
    \While{not converged}  
	{
        \space $logits_{teacher}$ $\leftarrow$ model\_teacher $(x_{full})$
        \\
        $logits$ $\leftarrow$ model $(x_{visible})$
        \\
        loss  $\leftarrow$ $DICE\left( {logits,target} \right) + w \times D_\alpha ^H\left( {\frac{{logits}}{\tau },\frac{{logit{s_{teacher}}}}{\tau }} \right)$
        
                
	}
 {\bfseries Output:} segmentation map.\\
\end{algorithm}

\section{Experiments}

\subsection{Dataset and Evaluation Metric}
\subsubsection{Dataset}
In this section, we employ the dataset from the Multimodal Brain Tumor Segmentation Challenge (BraTS) \cite{brats}, specifically BraTS 2018 and BraTS2020. BraTS 2018 comprises 285 cases with publicly available ground truth annotations. The BraTS datasets consist of multi-contrast MRI exams, encompassing four sequences: FLAIR, T1, T1c, and T2. The BraTS2020 dataset encompasses MRI scans of brain tumors meticulously annotated with ground truth segmentations. It encompasses a diverse array of brain tumor types, including gliomas, and offers multimodal imaging data, incorporating T1-weighted, T1-weighted with contrast enhancement, T2-weighted, and FLAIR MRI sequences.

\subsubsection{Evaluation Metric}
The Dice coefficient \cite{diceloss} is employed to measure the segmentation performance of the proposed method, defined as:
\begin{equation}
\label{equ_15}
    \operatorname{Dice}_{\bar{k}}(\hat{y}, y)=\frac{2 \cdot\left\|\hat{y}_{\bar{k}} \bigcap y_{\bar{k}}\right\|_1}{\left\|\hat{y}_{\bar{k}}\right\|_1+\left\|y_{\bar{k}}\right\|_1},
\end{equation}
where $\bar{k}$ represents distinct tumor categories, including BG, NCR/NE, ED, and ET. The entirety of the tumor, along with its core and enhancing region, comprises various combinations. Dice score $_{\bar{k}}$ denotes the similarity score of tumor category $\bar{k}$. Higher Dice scores signify predictions that closely align with the ground truth, thereby indicating superior segmentation accuracy.
\renewcommand\arraystretch{1.0}
\begin{table*}[!t]
 \setlength{\belowdisplayskip}{0pt}
 \setlength{\abovedisplayskip}{0pt}
 \setlength{\abovecaptionskip}{0pt}
 \centering
 \scriptsize
 \caption{Comparison of different methods (Dice \%) for various missing scenarios on the BRATS 2018 dataset where ◦ and • denote the missing and available modalities, respectively; bold denotes the best results.}
 \setlength{\tabcolsep}{1.2pt}
 \begin{tabular}{p{1.8cm}|p{1.0cm}p{1.0cm}p{1.0cm}p{1.0cm}p{1.0cm}|
 p{1.0cm}p{1.0cm}p{1.0cm}p{1.0cm}p{1.0cm}|
 p{1.0cm}p{1.0cm}p{1.0cm}p{1.0cm}p{1.0cm}}  
  \toprule[2pt] 
  \bf 
  Using Modality &\multicolumn{5}{c|}{\bf Whole Tumor}&\multicolumn{5}{c|}{\bf Tumor Core}&\multicolumn{5}{c}{ \bf Enhancing Tumor}\\
  \midrule[1.0pt]
  {F \space T1 \space T1c  T2}& MA3E \cite{M3AE}&MTI \space\space \cite{Multimodal}&D2-Net\cite{D2-Net}&RF-Net \cite{rfnet}
  &Our& MA3E\cite{M3AE}&MTI \space\space\cite{Multimodal}&D2-Net\cite{D2-Net}&RF-Net\cite{rfnet} &Our& MA3E\cite{M3AE}&MTI \space\space\cite{Multimodal}&D2-Net\cite{D2-Net}&RF-Net \cite{rfnet} &Our\\
  \midrule[1.0pt]
  {$\circ$ \space\space $\circ$ \space\space $\circ$ \space\space $\bullet$}
  &84.8&86.5&76.3&85.1&\bf88.4
  &69.4&68.7&56.7&66.9&\bf73.6
  &47.6&41.4&16.0&43.0&\bf51.3\\
  {$\circ$ \space\space $\circ$ \space\space $\bullet$ \space\space $\circ$}
  &75.8&77.7&42.8&73.6&\bf78.5
  &82.9&81.5&65.1&80.3&\bf83.0
  &73.7&75.7&66.3&67.7&\bf79.1\\
  {$\circ$ \space\space $\bullet$ \space\space $\circ$ \space\space $\circ$}
  &74.4&78.6&15.5&74.8&\bf79.8
  &66.1&65.5&16.8&65.2&\bf69.4
  &37.1&44.4&8.1&32.3&\bf45.1\\
  {$\bullet$ \space\space $\circ$ \space\space $\circ$ \space\space $\circ$}
  &88.7&88.4&84.1&85.8&\textbf{89.4}
  &66.4&66.7&47.3&62.6&\textbf{70.2}
  &35.6&40.5&8.1 &35.5&\textbf{46.5}\\
  {$\circ$ \space\space $\circ$ \space\space $\bullet$ \space\space $\bullet$}
  &86.3&88.2&84.1&85.6&\bf89.0
  &84.2&84.7&80.3&82.4&\bf84.8
  &75.3&77.7&68.7&70.6&\bf81.1\\
  {$\circ$ \space\space $\bullet$ \space\space $\bullet$ \space\space $\circ$}
  &77.2&81.8&62.1&77.5&\bf82.3
  &83.4&\bf83.5&78.2&81.3&81.4
  &74.7&77.1&70.7&68.5&\bf79.7\\
  {$\bullet$ \space\space $\bullet$ \space\space $\circ$ \space\space $\circ$}
  &89.0&89.7&87.3&89.0&\bf90.0
  &70.8&71.9&62.6&72.2&\bf72.6
  &41.2&44.4&17.4&38.5&\bf49.3
\\
  {$\circ$ \space\space $\bullet$ \space\space $\circ$ \space\space $\bullet$}
  &86.7&88.1&80.1&85.4&\bf88.9
  &71.8&72.3&63.2&71.1&\bf72.9
  &48.7&47.7&16.5&42.9&\bf50.0
\\
  {$\bullet$ \space\space $\circ$ \space\space $\circ$ \space\space $\bullet$}
  &89.9&90.2&87.9&89.3&\bf90.2
  &70.9&71.8&62.6&71.8&\bf72.0
  &45.4&48.2&17.4&45.4&\bf48.5\\
  {$\bullet$ \space\space $\circ$ \space\space $\bullet$ \space\space $\circ$}
  &89.7&89.4&87.5&89.4&\bf89.7
  &84.4&\bf84.8&80.8&81.6&82.6
  &75.0&76.8&64.8&72.5&\bf80.0
\\
  {$\bullet$ \space\space $\bullet$ \space\space $\bullet$ \space\space $\circ$}
  &88.9&90.3&87.7&89.9&\bf90.3
  &84.1&\bf85.1&80.9&82.3&84.2
  &74.0&77.3&65.7&71.1&\bf81.3
\\
  {$\bullet$ \space\space $\bullet$ \space\space $\circ$ \space\space $\bullet$}
  &89.9&89.7&88.4&90.0&\bf90.6
  &72.7&\bf74.0&63.7&74.0&73.0
  &44.8&50.0&19.4&46.0&\bf53.5\\
  {$\bullet$ \space\space $\circ$ \space\space $\bullet$ \space\space $\bullet$}
  &90.2&\bf90.5&88.8&90.4&89.6
  &84.6&\bf85.7&80.7&82.6&84.2
  &73.8&76.5&66.4&73.1&\bf80.2\\
  {$\circ$ \space\space $\bullet$ \space\space $\bullet$ \space\space $\bullet$}
  &85.7&\bf88.3&80.9&86.1&88.0
  &84.4&\bf85.8&79.0&82.9&83.5
  &75.4&78.4&68.3&70.9&\bf81.3\\
  {$\bullet$ \space\space $\bullet$ \space\space $\bullet$ \space\space $\bullet$}
  &90.1&90.5&88.8&\bf90.6&90.3
  &84.5&\bf85.9&80.1&82.9&84.0
  &75.5&80.3&68.4&71.4&\bf81.3\\
  \midrule[1.0pt]\centering
    {Average}
  &85.8&87.2&76.1&85.5&\bf87.6
  &77.4&77.8&66.5&75.6&\bf78.0
  &59.9&62.4&42.8&56.6&\bf65.8
\\
\bottomrule[2pt]
 \end{tabular}
 \label{main_res}
\end{table*}


\begin{table*}[!t]
 \setlength{\belowdisplayskip}{0pt}
 \setlength{\abovedisplayskip}{0pt}
 \setlength{\abovecaptionskip}{0pt}
 \centering
 \scriptsize
 \caption{Comparison of different methods (Dice \%) for various missing scenarios on the BRATS 2020 dataset where ◦ and • denote the missing and available modalities, respectively; bold denotes the best results.}
 \setlength{\tabcolsep}{1.2pt}
 \begin{tabular}{p{1.8cm}|p{1.0cm}p{1.0cm}p{1.0cm}p{1.0cm}p{1.0cm}|
 p{1.0cm}p{1.0cm}p{1.0cm}p{1.0cm}p{1.0cm}|
 p{1.0cm}p{1.0cm}p{1.0cm}p{1.0cm}p{1.0cm}}  
  \toprule[2pt] 
  \bf 
  Using Modality &\multicolumn{5}{c|}{\bf Whole Tumor}&\multicolumn{5}{c|}{\bf Tumor Core}&\multicolumn{5}{c}{ \bf Enhancing Tumor}\\
  \midrule[1.0pt]
  {F \space T1 \space T1c  T2}&
  U-HVED \cite{U-HVED} &MTI \space\space \cite{Multimodal}&mmFormer \cite{mmFormer}&RF-Net \cite{rfnet}
  &Our
  & U-HVED \cite{U-HVED} &MTI \space\space \cite{Multimodal}&mmFormer \cite{mmFormer}&RF-Net \cite{rfnet}
  &Our
  & U-HVED \cite{U-HVED} &MTI \space\space \cite{Multimodal}&mmFormer \cite{mmFormer}&RF-Net \cite{rfnet}
  &Our\\
  \midrule[1.0pt]
  {$\circ$ \space\space $\circ$ \space\space $\circ$ \space\space $\bullet$}
  &80.7&86.4&85.5&86.0&\bf87.1
  &57.4&71.4&63.3&71.0&\bf71.5
  &28.7&45.5&49.0&46.2&\bf52.7\\
  {$\circ$ \space\space $\circ$ \space\space $\bullet$ \space\space $\circ$}
  &68.5&77.3&78.0&76.7&\bf78.4
  &73.0&83.3&81.5&81.5&\bf84.1
  &66.6&78.9&78.3&74.8&\bf82.0
  \\
  {$\circ$ \space\space $\bullet$ \space\space $\circ$ \space\space $\circ$}
  &54.9&78.0&76.2&77.1&\bf78.1
  &36.7&66.8&63.2&66.0&\bf66.9
  &12.3&41.2&37.6&37.3&\bf45.9\\
  {$\bullet$ \space\space $\circ$ \space\space $\circ$ \space\space $\circ$}
  &82.6&89.0&86.5&87.3&\bf89.5
  &51.1&69.2& 64.6&69.1&\bf69.3
  &20.8&43.6&36.6 &38.1&\bf49.5\\
  {$\circ$ \space\space $\circ$ \space\space $\bullet$ \space\space $\bullet$}
  &83.3&88.3&87.5&87.7&\bf88.4
  &77.8&\bf86.3&82.6&83.4&84.6
  &68.7&81.6&77.2&75.9&\bf85.3
  \\
  {$\circ$ \space\space $\bullet$ \space\space $\bullet$ \space\space $\circ$}
  &71.5&81.1& 80.7&81.2&\bf82.0
  &76.4&85.2& 82.8&83.4&\bf85.2
  &67.8&79.2&81.7&75.9&\bf84.2
  \\
  {$\bullet$ \space\space $\bullet$ \space\space $\circ$ \space\space $\circ$}
  &85.0&89.9&88.7&89.7&\bf89.9
  &55.1&73.9& 71.7&73.0&72.5
  &22.5&48.1&42.9&40.9&\bf51.2
\\
  {$\circ$ \space\space $\bullet$ \space\space $\circ$ \space\space $\bullet$}
  &81.5&88.0& 86.9&87.7&\bf88.3
  &65.3&\bf73.2& 67.7&73.1&72.0
  &28.7&50.0&49.1&45.6&\bf52.7
\\
  {$\bullet$ \space\space $\circ$ \space\space $\circ$ \space\space $\bullet$}
  &87.4&90.5&89.4&89.8&\bf90.6
  &61.8&75.4&70.3&74.1&71.1
  &30.4&48.6& 49.0&49.3&\bf54.1
  \\
  {$\bullet$ \space\space $\circ$ \space\space $\bullet$ \space\space $\circ$}
  &86.1&89.9&89.3&89.8&\bf90.0
  &76.8&85.4&83.7&84.6&\bf85.4
  &69.5&81.7& 79.4&76.6&\bf84.0
\\
  {$\bullet$ \space\space $\bullet$ \space\space $\bullet$ \space\space $\circ$}
  &87.1&90.6&89.7&90.6&90.6
  &79.5&86.5&84.4&85.0&85.3
  &71.3&81.8& 80.6&76.8&\bf82.3
\\
  {$\bullet$ \space\space $\bullet$ \space\space $\circ$ \space\space $\bullet$}
  &88.0&90.3& 89.8&90.6&\bf90.6
  &63.4&75.9& 72.4&75.1&73.4
  &30.6&52.5& 50.0&49.9&\bf55.2
  \\
  {$\bullet$ \space\space $\circ$ \space\space $\bullet$ \space\space $\bullet$}
  &88.3&\bf90.6&90.4&90.6&89.9
  &78.6&\bf86.3&83.9&84.9&85.0
  &69.8&80.9& 78.7&77.1&\bf83.3
  \\
  {$\circ$ \space\space $\bullet$ \space\space $\bullet$ \space\space $\bullet$}
  &84.2&88.7&87.6&88.2&\bf89.5
  &79.9&\bf86.4&79.0&83.6&85.5
  &69.7&78.4&68.3&77.3&\bf81.2
  \\
  {$\bullet$ \space\space $\bullet$ \space\space $\bullet$ \space\space $\bullet$}
  &88.8&90.6&90.5&\bf91.1&90.7
  &80.4&\bf87.4&84.6&85.2&85.2
  &70.5&81.5&79.9&78.0&\bf85.0\\
  \midrule[1.0pt]\centering
    {Average}
  &81.2&87.3&76.1&86.9&-
  &67.1&79.5&66.5&78.2&-
  &48.5&69.9&42.8&61.4&-
\\
\bottomrule[2pt]
 \end{tabular}
 \label{main_res_2020}
\end{table*}

\subsection{Implementation Details}
For BraTS2018, we partition each dataset into training and validation sets, adhering to the same configuration as RFNet \cite{rfnet}. As per the challenge guidelines, the four intra-tumor structures (edema, enhancing tumor, necrotic core, and non-enhancing tumor core) are amalgamated into three tumor regions for evaluation: 1. The whole tumor, encompassing all tumor tissues. 2. The tumor core, comprising the enhancing tumor, necrotic core, and non-enhancing tumor core. 3. The enhancing tumor. 

Our model is trained on four Nvidia GTX3090 GPUs with 24GB of GPU memory. Following previous works \cite{wang2023SwinMM}, we divided the 3D images into sub-volumes of size $128 \times 128 \times 128$ and randomly masked a portion of them during pre-training. We train the model using the AdamW \cite{pp200} optimizer with a warm-up cosine scheduler for 50 epochs, an initial learning rate of $1 \times e^{-4}$, momentum of 0.99, and decay of $1 \times e^{-5}$, respectively. The total training epoch is 800, with 300 epochs in pre-training and fine-tuning, respectively. Our model is implemented in PyTorch \cite{p7} and MONAI \cite{p8}. For inference on these datasets, we applied double slicing window inference, where the window size is $64 \times 64 \times 64$ and the overlap between windows is 50\%.

\subsection{Comparison with the State-of-the-Art Methods}
\subsubsection{Compare Experimental Models}
The experimental models employ in this study for comparison include RF-Net \cite{rfnet},  MMFormer \cite{mmFormer}, D2-Net \cite{D2-Net}, MA3E \cite{M3AE}, and MTI \cite{Multimodal}. Each of these models contributes uniquely to the field of addressing incomplete data encountered in multi-modal brain tumor segmentation, with their specific attributes and advancements delineated as follows:

\begin{itemize}
              \item RFNet (\textit{Ding et al., CVPR, 2021})  \cite{rfnet}: RFNet is an innovative network designto tackle the challenge of incomplete data encountered in multi-modal brain tumor segmentation.

              \item MMFormer (\textit{Zang et al., MICCAI, 2022}) \cite{mmFormer}: MMFormer is an innovative model specifically crafted to address the complex challenges of incomplete multimodal learning in the realm of brain tumor segmentation within image processing. 
              
              \item D2-Net  (\textit{Yang et al., IEEE TMI, 2022})  \cite{D2-Net}: D2-Net introduces a novel dual disentanglement network tailored to facilitate brain tumor segmentation, particularly in scenarios where certain modalities are unavailable.
              
              \item MA3E (\textit{Liu et al., AAAI, 2023}) \cite{M3AE}: M3AE presents a pioneering approach specialized in segmenting brain tumors amidst the absence of specific modalities, placing particular emphasis on the pivotal role of multimodal representation learning. 

                \item MTI (\textit{Ting and Liu, JBHI, 2024}) \cite{Multimodal}:  
                MTI introduces a multimodal transformer model specifically designed to tackle brain tumor segmentation using incomplete MRI data. 
\end{itemize}

\subsubsection{Experimental Results and Analysis}
Table \ref{main_res} compares the performance of our model with other state-of-the-art models on the BraTS2018 and BraTS2020. The results of other models are taken from the original papers as they follow the same configuration as RFNet \cite{rfnet}. We select the following models for comparison: M$^3$AE \cite{M3AE}, a comprehensive model comprising pre-training and fine-tuning stages; MTI \cite{Multimodal}, a model utilizing separate modality encoders for feature capture and fusion; D$^2$-Net, a dual disentanglement network consisting of modality and tumor-region disentanglement stages; and RFNet \cite{rfnet}, featuring a region-aware fusion module for aggregating multi-modal features.

Our model achieves the best performance in most missing modality situations, especially when more than one modality is missing, where it significantly improves Dice performance by 5\% to 10\%. Particularly noteworthy is the enhanced predictive accuracy of the baseline for enhancing tumors, greatly improved by our model. For instance, when only T1 and T1c modalities are available, our model achieves almost the same accuracy as the full modality model in enhancing tumor segmentation. Moreover, our model improves the performance of existing state-of-the-art algorithms by an average of 1.35\% in Dice. In conclusion, these results underscore the effectiveness of our method, particularly in scenarios with a relatively large number of missing modalities.

\subsection{Ablation Study}
In this section, we elucidate the enhancement and efficacy stemming from each constituent of our training framework. We present the Dice scores for WT, TC, and ET, along with their mean values. All models undergo a pre-training phase consisting of 800 epochs followed by fine-tuning for an additional 300 epochs.

\subsubsection{Mask Ratio}
\begin{figure}[ht!]
    \centering
    \includegraphics[width=8cm]{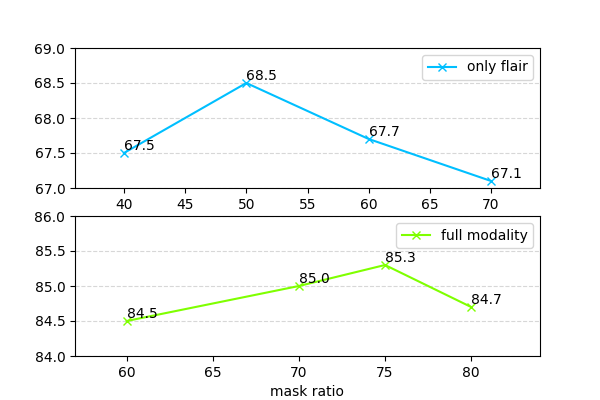}
    \caption{Ablation results showing the impact of different \textbf{mask ratios} when using only FLAIR and full modality.}
    \label{fig:maskratio}
\end{figure}

In order to determine the parameters $k$ and $b$ in Eq. \eqref{equ_6}, we conducted experiments under two scenarios: when three modalities are missing ($m = 3$) and when no modality is missing ($m = 0$). The impact of the masking ratio is illustrated in Fig. \ref{fig:maskratio}. For the case of missing three modalities, an optimal ratio of $50\%$ is observed. This result is intuitive, as excessively masking the input makes the task of reconstructing both the masked patches and the missing modality overly challenging. Conversely, in the scenario with no missing modalities, a higher mask ratio ($75\%$) is applied, simplifying the task to solely reconstructing the masked patches.

In our experiments, we utilize masking ratios of $0.5$, $0.6$, $0.65$, and $0.75$ for scenarios with missing three, two, one, and no missing modalities, respectively. This choice aligns with conclusions drawn in previous studies \cite{MAE,SimMIM}, indicating that a higher masking ratio is optimal for image data due to information redundancy.

\subsubsection{Reconstruction Target}
We conduct a comparison of different reconstruction targets when utilizing only the Flair modality. The results are summarized in Table \ref{rec_target}. We examined two permutations: whether random patches are masked during pre-training and whether the missing modalities are predicted. When only random patches are masked, the approach resembles the original SimMIM \cite{SimMIM} in pre-training. This method displayed some improvement compared to no masking, indicating the effectiveness of masked image modeling even with only one visible modality.

Alternatively, if only predicting the missing modality without masking, the model resembles a variational autoencoder (VAE) \cite{pp201}. However, at this stage, the model appears to have learned little during pre-training, resulting in a significant decrease in performance.

\begin{table}[!h]
 \setlength{\belowdisplayskip}{0pt}
 \setlength{\abovedisplayskip}{0pt}
 \setlength{\abovecaptionskip}{0pt}
 \centering
 \scriptsize
  \caption{Ablation results of different \textbf{reconstruction targets}, where “mask” indicates masking random patches during pre-training with a mask ratio of 0.5, and “predict” denotes the reconstruction target includes the missing modalities.
}
  \setlength{\tabcolsep}{1.2pt}
\begin{tabular}{>{\centering\arraybackslash}p{1.5cm}>{\centering\arraybackslash}p{1.5cm}|>{\centering\arraybackslash}p{1.5cm}>{\centering\arraybackslash}p{1.5cm}>{\centering\arraybackslash}p{1.5cm}>{\centering\arraybackslash}p{1.5cm}}
\toprule[1.5pt]
\multicolumn{2}{c|}{Pretrain} &
\multicolumn{4}{c}{Dice} \\
\midrule[1.0pt]
 \multicolumn{1}{c|}{Mask} & Predict
   &
  \multicolumn{1}{c|}{WT} & 
  \multicolumn{1}{c|}{TC} &
  \multicolumn{1}{c|}{ET} & Mean
 \\ \midrule[1.0pt]
  \multicolumn{1}{c|}{$\times$} & $\times$
   &
  \multicolumn{1}{c|}{87.6} & 
  \multicolumn{1}{c|}{66.8} &
  \multicolumn{1}{c|}{43.8} & 66.1
   \\ \midrule[1.0pt]
  \multicolumn{1}{c|}{\checkmark} &$\times$
   &
  \multicolumn{1}{c|}{88.2\color[HTML]{009901}(+0.6)} &
  \multicolumn{1}{c|}{68.2\color[HTML]{009901}(+1.4)} &
  \multicolumn{1}{c|}{45.0\color[HTML]{009901}(+1.2)} & 67.1\color[HTML]{009901}(+1.0)
   \\ \midrule[1.0pt]

  \multicolumn{1}{c|}{$\times$} &\checkmark
   &
  \multicolumn{1}{c|}{70.6\color[HTML]{FE0000}(-17.0)} &
  \multicolumn{1}{c|}{50.8\color[HTML]{FE0000}(-16.0)} &
  \multicolumn{1}{c|}{39.6\color[HTML]{FE0000}(-4.2)} & 53.7\color[HTML]{FE0000}(-12.4)
   \\ \midrule[1.0pt]
  \multicolumn{1}{c|}{\checkmark} & \checkmark
   &
  \multicolumn{1}{c|}{89.4\color[HTML]{009901}(+1.9)} &
  \multicolumn{1}{c|}{70.2\color[HTML]{009901}(+3.4)} &
  \multicolumn{1}{c|}{46.5\color[HTML]{009901}(+2.7)} & \bf68.7\color[HTML]{009901}(+2.6)
   \\ \bottomrule[2.0pt]
\label{rec_target}
\end{tabular}
\end{table}
However, when we combine the two reconstruction targets as describe in the preceding section, the performance improves significantly. This outcome underscores the effectiveness of our approach.
\subsubsection{Hölder Conjugate Exponents}

Table \ref{holder_table} presents a comparative analysis of performance across various Hölder parameters ($\alpha$), considering both KLD and no usage of knowledge distillation. Through our analyses, it is evident that employing Hölder divergence with $\alpha = 1.6$ yields substantial enhancement, with a slight improvement of 0.5 compared to no usage of knowledge distillation. These findings underscore the significance of selecting appropriate Hölder conjugate exponents ($\alpha$) to achieve notable performance boosts. Additionally, it is noteworthy that when $\alpha = 2.0$, the Hölder divergence is equivalent to Cauchy-Schwarz divergence, resulting in performance similar to KLD.

\begin{table}[!h]
 \setlength{\belowdisplayskip}{0pt}
 \setlength{\abovedisplayskip}{0pt}
 \setlength{\abovecaptionskip}{0pt}
 \centering
 \scriptsize
  \caption{Ablation results of different \textbf{Hölder conjugate exponents}, compared with KL divergence when only using modality T2.}
  \setlength{\tabcolsep}{8pt}
\begin{tabular}{>{\centering\arraybackslash}p{1.0cm}>{\centering\arraybackslash}p{1.0cm}|>{\centering\arraybackslash}p{3cm}>{\centering\arraybackslash}p{3cm}>{\centering\arraybackslash}p{3cm}>{\centering\arraybackslash}p{1.0cm}}
\toprule[1.5pt]
\multicolumn{2}{c|}{Fine Tune} &
\multicolumn{4}{c}{Dice} \\
\midrule[1.0pt]
 \multicolumn{1}{c|}{divergence} & $\alpha$
   &
  \multicolumn{1}{c|}{WT} & 
  \multicolumn{1}{c|}{TC} &
  \multicolumn{1}{c|}{ET} & Mean
  \\ \midrule[1.0pt]
  \multicolumn{1}{c|}{-} & -
   &
  \multicolumn{1}{c|}{87.5} & 
  \multicolumn{1}{c|}{72.0} &
  \multicolumn{1}{c|}{50.2} & 69.9
 \\ \midrule[1.0pt]
  \multicolumn{1}{c|}{KL} & -
   &
  \multicolumn{1}{c|}{88.0} & 
  \multicolumn{1}{c|}{72.8} &
  \multicolumn{1}{c|}{50.5} & 70.4
   \\ \midrule[1.0pt]
  \multicolumn{1}{c|}{Hölder} & 1.5
   &
  \multicolumn{1}{c|}{88.4} &
  \multicolumn{1}{c|}{73.2} &
  \multicolumn{1}{c|}{51.5} & 71.0
   \\ \midrule[1.0pt]

  \multicolumn{1}{c|}{\bf Hölder} & \bf1.6
   &
  \multicolumn{1}{c|}{\bf88.4} &
  \multicolumn{1}{c|}{\bf73.6} &
  \multicolumn{1}{c|}{\bf51.3} & \bf71.1
   \\ \midrule[1.0pt]
  \multicolumn{1}{c|}{Hölder} & 1.7
   &
  \multicolumn{1}{c|}{88.3} &
  \multicolumn{1}{c|}{72.8} &
  \multicolumn{1}{c|}{51.1} & 70.7
  \\ \midrule[1.0pt]
  \multicolumn{1}{c|}{Hölder} & 1.8
   &
  \multicolumn{1}{c|}{88.2} &
  \multicolumn{1}{c|}{72.5} &
  \multicolumn{1}{c|}{50.4} & 70.4
  \\ \midrule[1.0pt]
  \multicolumn{1}{c|}{Hölder} & 2.0
   &
  \multicolumn{1}{c|}{88.2} &
  \multicolumn{1}{c|}{72.7} &
  \multicolumn{1}{c|}{50.6} & 70.5
   \\ \bottomrule[2.0pt]
\label{holder_table}
\end{tabular}
\end{table}

Overall, our findings underscore the effectiveness and significance of Hölder divergence (HD) in improving performance in knowledge distillation for segmentation tasks. Throughout our experiments, we consistently observed enhanced results by setting the Hölder Conjugate Exponent $\alpha$ to 1.6, reaffirming its efficacy.

\subsection{Discussion and Limitations}
Our proposed method, which combines masked predicted auto-encoder pre-training with Hölder divergence-based knowledge distillation, showcase significant advancements in handling missing modalities in brain tumor segmentation. By employing masked predicted auto-encoders, our model effectively learns robust features from incomplete data, while the integration of Hölder divergence in the knowledge distillation process enhances the transfer of knowledge from complete modality networks to those with missing modalities.

The adoption of Hölder divergence over traditional KLD demonstrate a more balanced distribution of attention across different tumor categories. This proves particularly advantageous when certain modalities, emphasizing specific features, are absent. As a result, our approach yields more accurate and balanced segmentations across various tumor regions, even in scenarios where certain modalities are missing.

Our ablation study provide crucial insights into the contributions of each component within our framework. It highlights the importance of carefully choosing the mask ratio during pre-training and the significance of predicting both masked patches and missing modalities. Additionally, it underscores the impact of selecting appropriate Hölder conjugate exponents for optimal performance during the knowledge distillation phase.

Despite the promising results, our method has several limitations:
   \begin{itemize}

       \item Complexity of Model Training: Our model needs to be modified for each specific situation where modalities are missing, resulting in the need to train separate models for each scenario. This approach is computationally intensive, requiring the training of $2^N - 1$ models for N modalities.

        \item Sensitivity to Hyperparameters: The performance of our model is sensitive to the choice of hyperparameters, such as the mask ratio during pre-training and the Hölder conjugate exponents during knowledge distillation. Finding the optimal set of hyperparameters may necessitate extensive experimentation and validation.
   \end{itemize}

\section{Conclusion}
In this work, we introduce a novel approach to incomplete modality brain tumor segmentation, unveiling a self-supervised pre-training framework inspired by “predict next word” techniques and incorporating Hölder divergence-based knowledge distillation. Our comprehensive experiments conducted on the BRATS2018 and BRATS2020 demonstrate remarkable performance, surpassing existing methods in various missing modality scenarios. The consistent enhancement observed in Dice scores across different tumor regions validates the efficacy of our approach in improving segmentation accuracy and robustness.

While our method exhibits promising results, there are avenues for future exploration. The computational complexity associated with training separate models for each missing modality scenario poses a significant challenge. Further research is warranted to address this issue and extend the applicability of our approach to other multi-modality datasets. Additionally, investigating strategies to mitigate the impact of missing modalities on segmentation accuracy remains an important direction for future work.

\newpage
\bibliographystyle{IEEEtran}
\bibliography{IEEETMI}

\end{document}